\documentclass[11pt]{article}
\usepackage[dvipdfmx]{graphicx}   
\usepackage{amssymb}
\usepackage{array} 
\usepackage[authoryear]{natbib}
\usepackage{bibentry}
\usepackage{amsmath}
\usepackage{float}
\usepackage{verbatim}
\usepackage{color}
\usepackage{makeidx}
\usepackage{bm}
\usepackage{appendix}
\usepackage{lscape}
\usepackage{subcaption}
\usepackage{rotating}
\usepackage{here}

\begin{document}
\title{Unified Mixture Sampler for State-Space Models: \\ 
Application to Stochastic Conditional Duration Models}
\author{\textsc{Daichi Hiraki} \\
%EndAName
\textit{Graduate School of Economics, University of Tokyo, Tokyo 113-0033, Japan} \\
\texttt{hdaichi397@gmail.com}
\and \textsc{Yasuhiro Omori} \\
%EndAName
\textit{Faculty of Economics, University of Tokyo, Tokyo 113-0033, Japan}\\
\texttt{omori@e.u-tokyo.ac.jp} \\
}
\date{\today}
\maketitle

\begin{abstract}
We propose a unified mixture sampler (UMS) that provides a universal estimation framework for nonlinear state-space models with `exp-exp' likelihood kernels. Unlike existing methods that require deriving new mixture approximations for each specific distribution, our approach dynamically adapts the standard ten-component mixture from Omori et al. (2007) through a deterministic re-centering and rescaling algorithm. Applying this to the stochastic conditional duration (SCD) model, we demonstrate that the proposed sampler can efficiently handle unknown shape parameters—such as those in Weibull or Gamma distributions—by updating mixture components near-instantaneously during MCMC iterations. The UMS not only simplifies implementation but also ensures exact inference via a lightweight Metropolis-Hastings step. Numerical examples show that our method substantially outperforms the conventional slice sampling approach, significantly reducing autocorrelation in MCMC samples while maintaining high computational efficiency. This unified framework encompasses a wide range of applications, including logit, Poisson, and various SCD model specifications, providing a highly efficient alternative to model-specific samplers.
\end{abstract}

\newpage
\section{Introduction}

State-space models, in which the latent state evolves over time and dictates the distribution of observable outcomes, are indispensable tools in financial and economic analysis. Despite their versatility, nonlinear and non-Gaussian state-space models pose significant inferential challenges, particularly regarding the efficient estimation of parameters and latent states. From a Bayesian perspective, which relies on Markov chain Monte Carlo (MCMC) algorithms, this difficulty primarily stems from the problem of constructing an effective proposal distribution for the latent variables.

A major breakthrough in addressing this challenge within the context of stochastic volatility models of \cite{Taylor(08)} was the introduction of the auxiliary mixture sampler by \cite{KimShephardChib(98)} and \cite{OmoriChibShephardNakajima(07)}. This approach involves two key steps: first, the observation equation is transformed into a linear state-space form; second, the non-Gaussian error term is approximated by a finite mixture of normal distributions, resulting in a conditionally linear Gaussian model. This transformation enables the application of highly efficient sampling techniques, such as the simulation smoother introduced by \cite{DeShephard(95)} and \cite{DurbinKoopman(02)}. Due to its computational power, this methodology has been extended to a wide variety of state-space models beyond the standard stochastic volatility framework.

However, several critical issues remain. First, the method is inherently model-specific, requiring a tailored mixture of normals for each distinct model. Second, the approach is only applicable if the model can be successfully linearized in its first step. Consequently, the auxiliary mixture sampler often lacks flexibility when considering model extensions. Third, if the likelihood approximated by the mixture depends on parameters that are updated during MCMC iterations, the mixture components must be re-optimized repeatedly, which is computationally prohibitive. A prominent example of a model facing these challenges is the stochastic conditional duration (SCD) model of \cite{BauwensVeredas(04)}. Specifically, when employing distributions such as the Weibull or Gamma for durations, the shape parameters must be estimated within the MCMC scheme. Since the distribution of the linearized error term changes with these parameters, the standard auxiliary mixture sampler would require re-calculating the optimal mixture constants at every iteration, which is practically infeasible.

In this paper, we address these limitations by proposing the unified mixture sampler (UMS), a versatile framework that efficiently applies the auxiliary mixture sampling principle to a broad class of nonlinear non-Gaussian state-space models. Our approach centers on the observation that many nonlinear models feature a likelihood component with an `exp-exp' structure, of the form $\exp(-A \exp(\cdot))$ with $A > 0$. We demonstrate that the UMS can be applied to this class by dynamically re-centering and rescaling the classic normal-mixture constants from the stochastic volatility literature. Crucially, our method bypasses the need for an initial linearization step and allows for the near-instantaneous update of mixture components, even when the likelihood contains time-varying shape parameters. Despite its computational simplicity, we confirm that this deterministic approach yields approximation accuracy comparable to model-specific optimizations. In our simulation experiments, we apply the UMS to the SCD model and find that it produces MCMC samples with significantly lower autocorrelation and requires substantially less execution time compared to existing methods such as the slice sampler.

\section{Auxiliary mixture sampler}
\label{sec:Auxiliary mixture sampler}
\subsection{Nonlinear state-space formulation}

We consider a general class of nonlinear state-space models where the observation equation is nonlinear with respect to the latent state variables:
\begin{align*}
    &y_t = f_\theta(h_t, \epsilon_t), \quad t=1,\dots,n, \\
    &h_{t+1} = \mu + \phi(h_t-\mu) + \eta_t, \quad \eta_t \sim N(0,\sigma^2), \quad t=1,\dots,n-1.
\end{align*}
The latent variables follow an autoregressive process of order one (AR(1)). In empirical studies of financial time series, the estimate of $\phi$ is usually close to one, resulting in the poor mixing of $h_t$ in MCMC algorithm. For simplicity, we temporarily assume that the error terms $\epsilon_t$ and $\eta_t$ are mutually independent and independent across time. 
Various established models fall into this class, each posing unique estimation challenges.\vspace{2mm}\\
\noindent
{\it Example 1}. Stochastic volatility (SV) model. It is  defined by
    \begin{equation}
        y_t = \exp(h_t/2) \epsilon_t, \quad \epsilon_t \sim N(0,1), 
    \end{equation}
    and its conditional likelihood is given by $p(y_t|h_t) = (2 \pi)^{-1/2} \exp\left\{ -h_t/2 - y_t^2 \exp(-h_t) /2 \right\}$. This includes `exp-exp' structure in terms of $h_t$.\vspace{2mm}\\
\noindent
{\it Example 2}. Stochastic conditional duration (SCD) model (see e.g. \cite{BauwensVeredas(04)}, \cite{StricklandForbesMartin(06)}, \cite{MenKolkiewiczWirjanto(15)}). Let us define as in SV model:
    \begin{equation}
        y_t = \exp(h_t/2) \epsilon_t, 
    \end{equation}
    where $\epsilon_t$ follows a distribution with positive support, such as the standardized exponential, Weibull, or Gamma distribution. It defines a class of SCD models where the corresponding densities are summarized in Table \ref{table:duration_densities}. Note that they also have `exp-exp' structure in terms of $h_t$. 
    \begin{table}[H]
    \centering
    \begin{tabular}{lll}
    \hline
    Distribution & $p(y_t | h_t, \theta)$ & $\theta$ \\ \hline
    Exponential & $\exp(-h_t) \exp \{ -y_t \exp(-h_t) \}$ & - \\
    Weibull & $\gamma \left( \frac{\Gamma(1+\gamma^{-1})}{\exp(h_t)} \right)^{\gamma} y_t^{\gamma-1} \exp \left\{ -\left[ \frac{y_t}{\exp(h_t)} \Gamma(1+\gamma^{-1}) \right]^{\gamma} \right\}$ & $\gamma$ \\
    Gamma & $\left( \frac{\zeta}{\exp(h_t)} \right)^{\zeta} y_t^{\zeta-1} \exp \left\{ -\frac{y_t \zeta}{\exp(h_t)} \right\} / \Gamma(\zeta)$ & $\zeta$ \\ \hline
    \end{tabular}
    \caption{Densities of SCD models (see \cite{StricklandForbesMartin(06)}).}
    \label{table:duration_densities}
    \end{table}
    These are state-space frameworks designed to characterize the dynamic evolution of time intervals between consecutive financial events by assuming a latent process drives the conditional mean of the durations.\vspace{2mm}\\
    \noindent
    {\it Example 3}. Type I extreme value distribution (the distribution of $-\log X$ where $X$ follows exponential distribution with mean 1). 
    %discrete response and count data models: 
    For example, \cite{FruhwirthFruhwirth(07)} and \cite{FruhwirthFruhwirthHeldRue(09)} consider logistic and time-varying Poisson models, where the likelihood involves Type I extreme value distribution. Moreover, it is used in the time series model of extreme values such as the max stable process (see e.g. \cite{KunihamaOmoriZhang(12)}, \cite{NakajimaKunihamaOmoriSchnatter(12)}).\vspace{2mm}\\
Such nonlinear observation equations are typically transformed into a linear state-space form to facilitate estimation. For instance, in the standard SV model, the transformation $\log y_t^2 = h_t + \log \epsilon_t^2$ linearizes the relationship with respect to $h_t$. However, the resulting error term follows a $\log \chi^2$ distribution, whose non-Gaussianity precludes the direct use of efficient Gaussian-based sampling techniques like the simulation smoother (\cite{DeShephard(95)}, \cite{DurbinKoopman(02)}). 

More generally, the likelihoods of these models often feature a kernel of the form $\exp(-A \exp(\cdot))$ for some $A > 0$. Such a structure typically arises when (i) the conditional distribution belongs to the exponential family, (ii) it is characterized by a positive parameter, and (iii) this parameter is parameterized as $\exp(h_t)$ to ensure its positivity. While most of these models can be handled via logarithmic transformations, they still pose a significant challenge for the standard auxiliary mixture sampler when the likelihood depends on unknown shape parameters (such as $\gamma$ in Weibull or $\zeta$ in Gamma distributions). In such cases, the mixture components must be re-optimized at each MCMC iteration, which is computationally prohibitive.

\subsection{Unified mixture approximation}
\label{subsec:unified_approximation}

To restore the conditionally linear Gaussian state-space structure, our approach builds upon the high-precision approximation of the $\log \chi^2$ density using a finite mixture of normal distributions:
\begin{equation*}
    \frac{1}{\sqrt{2\pi}} \exp \left( \frac{x - \exp(x)}{2} \right) \approx \sum_{i=1}^{K} p_i v_i^{-1} \phi \left( \frac{x-m_i}{v_i} \right),
\end{equation*}
where $\phi(\cdot)$ denotes the standard normal density function. The constants $(p_i, m_i, v_i^2)$ for a ten-component mixture ($K=10$) obtained from \cite{OmoriChibShephardNakajima(07)} are reproduced in Table \ref{table:approx_chi2}.

\begin{table}[H]
    \small
  \centering
  \begin{tabular}{crrr}
    \hline 
    $i$ & $p_i$  & $m_i$ & $v_i^2$ \\
    \hline
    1 & 0.00609 & 1.92677 & 0.11265 \\
    2 & 0.04775 & 1.34744 & 0.17788 \\
    3 & 0.13057 & 0.73504 & 0.26768 \\
    4 & 0.20674 & 0.02266 & 0.40611 \\
    5 & 0.22715 & -0.85173 & 0.62699 \\
    6 & 0.18842 & -1.97278 & 0.98583 \\
    7 & 0.12047 & -3.46788 & 1.57469 \\
    8 & 0.05591 & -5.55246 & 2.54498 \\
    9 & 0.01575 & -8.68384 & 4.16591 \\
    10 & 0.00115 & -14.65000 & 7.33342 \\
    \hline
  \end{tabular}
  \caption{Selection of mixture constants $(p_i, m_i, v_i^2)$ from \cite{OmoriChibShephardNakajima(07)}.}
  \label{table:approx_chi2}
  \normalsize
\end{table}

Crucially, this approximation can be extended to a broader class of `exp-exp' density kernels. Consider a target distribution with a kernel of the form:
\begin{equation}
    f(x; a, b, c) = \exp \left( \frac{a}{2}cx - \frac{b}{2}\exp(cx)\right), \quad -\infty < x < \infty,
    \label{eq:exp_exp_kernel}
\end{equation}
where $a \in \mathbb{R}$, $b > 0$, and $c \neq 0$. For $c = 1$, the approximation is derived by rearranging the kernel to match the $\log \chi^2$ form:
\begin{align*}
    \exp \left( \frac{a}{2}x - \frac{b}{2}\exp(x)\right) &= \exp\left( \frac{a-1}{2}x - \frac{\log b}{2} \right) \exp \left( \frac{x+\log b - \exp(x+\log b)}{2} \right) \\
    &\approx \exp\left( \frac{a-1}{2}x - \frac{\log b}{2} \right) \sqrt{2\pi} \sum_{i=1}^K p_i v_i^{-1} \phi \left( \frac{x+\log b-m_i}{v_i} \right).
\end{align*}
By completing the square within the exponential terms of the Gaussian components, the general kernel in \eqref{eq:exp_exp_kernel} is shown to be approximately proportional to a new mixture of normals:
\begin{equation*}
    f(x; a, b, c) \approx C \sum_{i=1}^K \tilde{p}_i \tilde{v}_i^{-1} \phi \left( \frac{x - \tilde{m}_i}{\tilde{v}_i} \right), 
\end{equation*}
where $C = 1 / \sum_{i=1}^K \tilde{p}_i$ is a normalizing constant, and the dynamically re-centered and rescaled mixture components are given by:
\begin{align}
    \label{equ:mixture p}
    \tilde{p}_i &= \frac{p_i}{|c|} \exp \left( -\frac{\log b}{2} + \frac{(1-a)(\log b - m_i)}{2} + \frac{v_i^2}{8}(1-a)^2 \right), \\
    \label{equ:mixture v}
    \tilde{v}_i &= \frac{v_i}{|c|}, \\
    \label{equ:mixture m}
    \tilde{m}_i &= \frac{m_i - \log b - \frac{1-a}{2}v_i^2}{c}.
\end{align}

When $a = b = c = 1$, this framework collapses to the original approximation of the $\log \chi^2$ distribution. For the standard SV model, the likelihood kernel $\exp(-x/2 - y^2 \exp(-x)/2)$ corresponds to $a=1$, $b=y^2$, and $c=-1$. Under this specification, the likelihood is approximated by $\sum_{i=1}^K p_i v_i^{-1} \phi \left( (x - \log y^2 + m_i)/v_i \right)$, which implies the linear Gaussian observation equation $\log y^2 \sim \sum_{i=1}^K p_i \times N(x+m_i, v_i^2)$. This unified approach encompasses a wide variety of models, including those involving distributions listed in Table \ref{table:duration_densities}.

\begin{figure}
    \centering
    \includegraphics[width=0.9\linewidth]{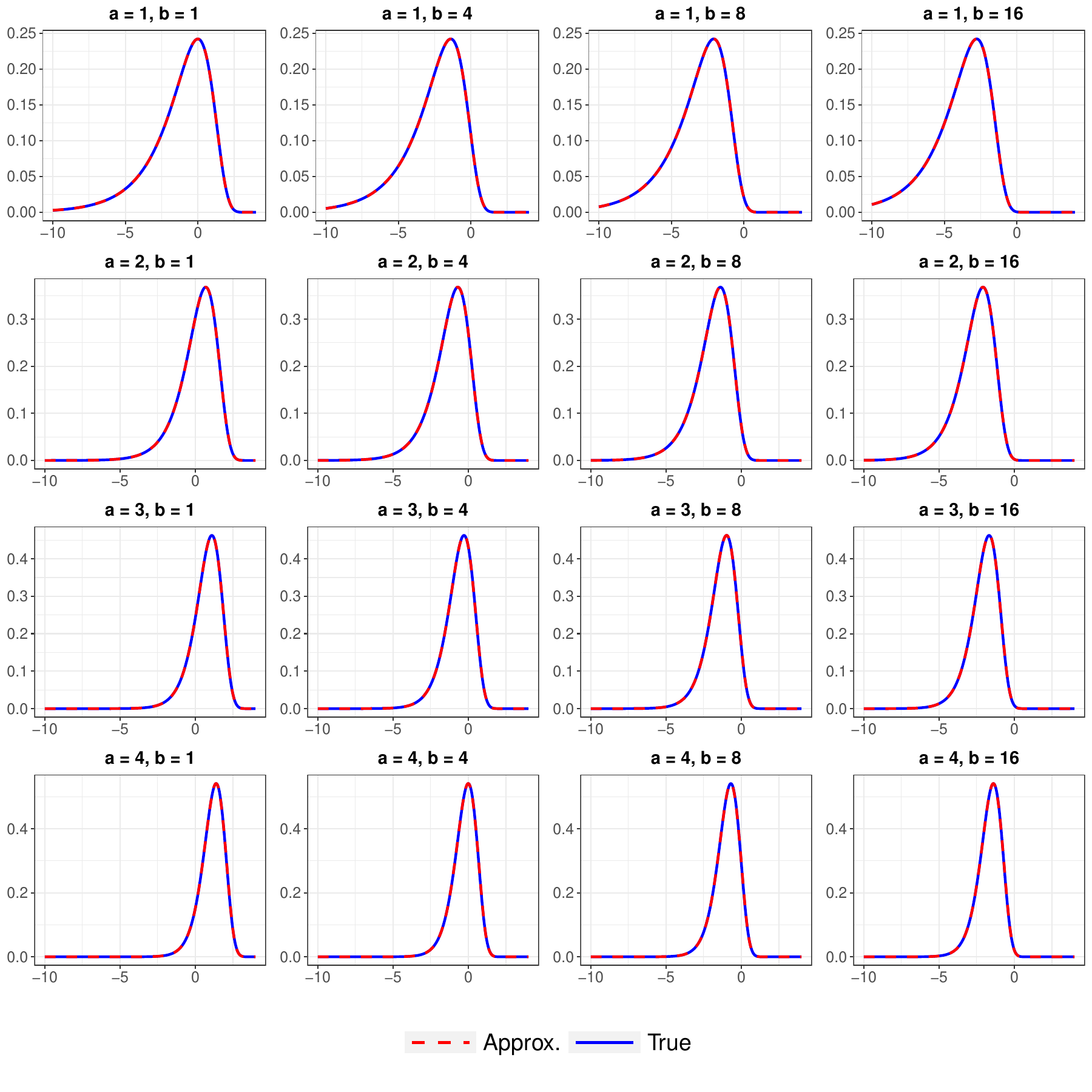}
    \caption{True density (blue) versus the unified mixture approximation (red) for $c = 1$.}
    \label{fig:approx}
\end{figure}

\begin{figure}
    \centering
    \includegraphics[width=0.9\linewidth]{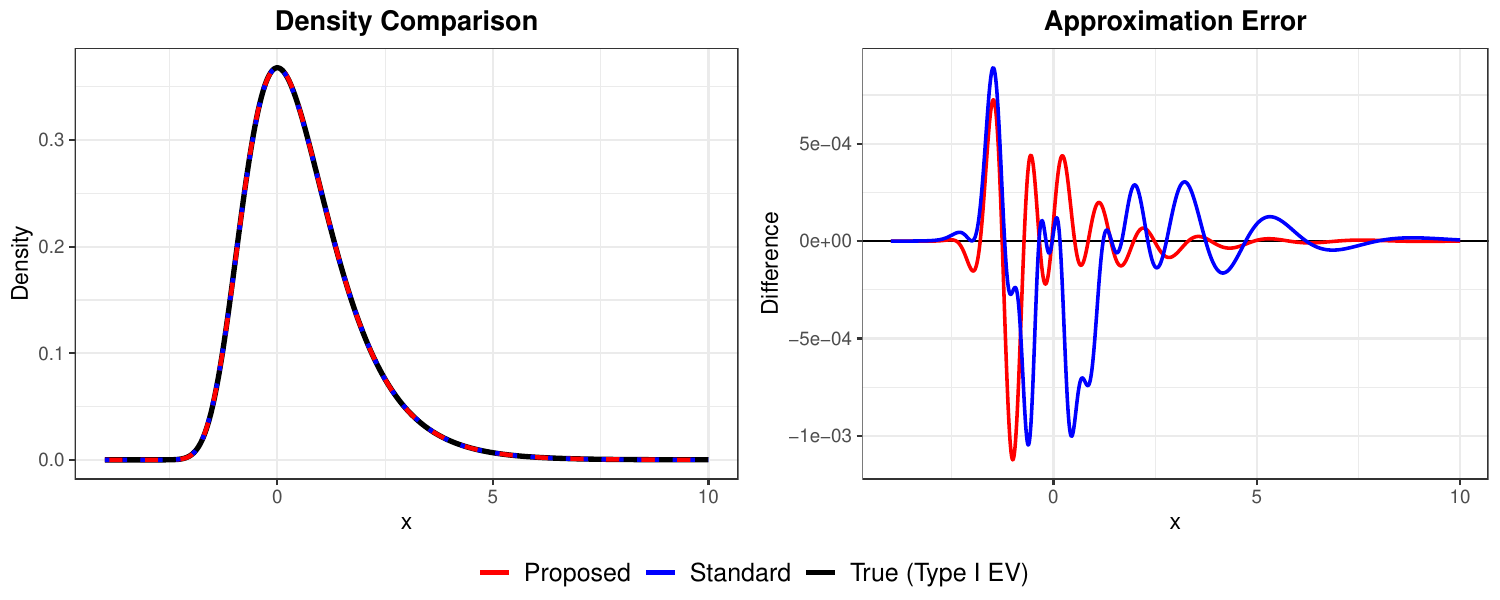}
    \caption{True density of Type I extreme value distribution (black) versus the unified mixture approximation (red) and the conventional mixture of normals from \cite{FruhwirthFruhwirth(07)} (blue). The right figure shows the difference between the true and the approximation curves.}
    \label{fig:comparison}
\end{figure}

As demonstrated by Figure \ref{fig:approx}, 
the unified mixture approximation closely overlaps with the true densities across various values of $(a, b)$. Furthermore, as shown in Figure \ref{fig:comparison}, our approach achieves approximation accuracy comparable to model-specific optimizations, such as those developed for the Type I extreme value distribution (\cite{FruhwirthFruhwirth(07)}).
The primary computational advantage of this unified sampler is that the updates for $(\tilde{p}_i, \tilde{v}_i, \tilde{m}_i)$ are purely deterministic and analytically tractable. This adds virtually no computational overhead even when the parameters $(a, b, c)$ are updated within MCMC iterations. This efficiency is particularly critical for models who have like the Weibull and Gamma distributions where optimization-based approximations would require computationally expensive re-evaluations at every iteration.


\newpage
\section{Application to the SCD model with Weibull distribution}
\label{sec:SCD model}
\subsection{Setup}
The density of conditional distribution for duration under Weibull distribution, $p(y_t|h_t, \gamma)$, given in Table \ref{table:duration_densities}, is
\begin{align*}
    p(y_t|h_t,\gamma) &= \gamma \left( \frac{\Gamma(1+\gamma^{-1})}{\exp(h_t)} \right)^{\gamma} y_t^{\gamma-1} \exp \left\{ -\left[ \frac{y_t}{\exp(h_t)} \Gamma(1+\gamma^{-1}) \right]^{\gamma} \right\}.
\end{align*}
As a function of $h_t$ conditional on ($y_t$, $\gamma$), this likelihood is proportional to 
  \begin{align*}
\exp \left\{ -\gamma h_t - y_t^{\gamma} \Gamma(1+\gamma^{-1})^{\gamma} \exp(-\gamma h_t) \right\}.
\end{align*}
Applying our unified framework to this kernel with parameters $a=2$, $b=2y_t^{\gamma} \Gamma(1+\gamma^{-1})^{\gamma}$, and $c=-\gamma$, we obtain the adapted mixture components ($\tilde{p}_i, \tilde{v}_j, \tilde{m}_j$) by equations \eqref{equ:mixture p}--\eqref{equ:mixture m}. That is $p(y_t|h_t, \gamma)$ is approximated by $\sum_{i=1}^K \tilde{p}_i \tilde{v}_i^{-1} \phi \left( (h_t - \tilde{m}_i)/\tilde{v}_i \right)$.
By introducing latent mixture indicators $s_t \in \{1,\dots,K\}$, the non-linear system is reduced to a linear Gaussian observation system:
\begin{equation*}
    \tilde{m}_{s_t} = h_t + \tilde{v}_{s_t} z_t, 
\end{equation*}
where $z_t \sim N(0,1)$. This formulation restores the standard linear Gaussian state-space form, enabling the use of the simulation smoother.

\subsection{MCMC algorithm}
Let $\theta = (\alpha,\gamma)$, where $\alpha = (\mu,\phi,\sigma^2)$ denotes the state equation parameters. The MCMC simulation proceeds in the following blocks:
\begin{itemize}
    \item[1.] Initialize the latent volatilities $h$ and the parameters $\theta=(\alpha,\gamma)$.
    \item[2.] Generate $\gamma|\alpha,h, y \sim \pi(\gamma|\alpha,h, y)$.
    \item[3.] Generate $(\alpha,h)|\gamma, y \sim \pi(\alpha,h|\gamma, y)$ jointly.
    \item[4.] Return to Step 2.
\end{itemize}
The details of the MCMC algorithm are as follows. 
\\
\noindent
{\it Step 2: Generation of $\gamma|\alpha,h,y$}. 
We use the random-walk Metropolis-Hastings method. For the current value $\gamma^o$, we generate a candidate value $\gamma^\dag$ from $\log(\gamma^\dag) \sim N(\log(\gamma^o), \nu^2)$ and decide whether to accept or reject it. In this paper, we set $\nu^2 = 0.1^2$.\vspace{2mm}\\
{\it Step 3: Joint generation of $(\alpha,h)|\gamma,y$}
\label{generate h}
To mitigate strong posterior correlations between the parameters and the latent states, we generate $(\alpha, h)$ jointly in a single block using a Metropolis-Hastings (MH) step. Since our re-centered mixture approximation is highly accurate, we utilize the approximate posterior distribution as an efficient proposal density, analogous to the strategy in \cite{ChibNardariShephard(02)}. 
Let $\pi(\alpha, h|\gamma, y) \propto f(y|h,\gamma) f(h|\alpha) \pi(\alpha)$ denote the exact conditional posterior. We define the approximate target density, augmented with the mixture indicators $s = (s_1, \dots, s_n)$, as:
\begin{align*}
    \pi^*(\alpha,h,s|\gamma,y) &= \pi^*(\alpha,h|\gamma,s,y) \times q(s) \\
         &= \pi^*(h|\alpha,\gamma,s,y)\pi^*(\alpha|\gamma,s,y) \times q(s),
\end{align*}
where $q(s) = \prod_{t=1}^n 
\tilde{p}_{s_{t}}$, and
\begin{align*}
    \pi^*(h|\alpha, \gamma, s, y) &= \frac{\prod_{t=1}^n g(\tilde{m}_{s_t}| h_t, \alpha, \gamma, s_t)}{m(\{\tilde{m}_{s_t}\}_{t=1,\ldots,n}|\alpha, \gamma, s)} \times f(h|\alpha), \\
    \pi^*(\alpha|\gamma, s, y) & \propto  m(\{\tilde{m}_{s_t}\}_{t=1,\ldots,n}|\alpha, \gamma, s) \pi(\alpha).
\end{align*}
Here, $g(\cdot | h_t, \dots)$ denotes the approximate Gaussian observation likelihood given by the mixture components, and $m(\cdot | \alpha, \gamma, s)$ is the normalizing constant (the marginal likelihood of the approximate linear state-space model), which can be efficiently evaluated using the Kalman filter. 

We generate a candidate draw $(\alpha^\dag, h^\dag, s)$ in two sub-steps and then apply an MH accept-reject step:
\begin{itemize}
    \item[(a)] \textit{Sample indicators $s$:} Generate $s \sim q(s|h,\alpha, \gamma, y)$, which is implemented by independently sampling $s_t$ and $s_{2t}$ with probabilities:
    \begin{equation*}
        Pr(s_t = i) \propto \tilde{p}_{i} f_N(\tilde{m}_i|h_t, \tilde{v}_{i}^2),
    \end{equation*}
    for $t = 1,\dots,n$, where $f_N(\cdot | m, s^2)$ is the density of $N(m, s^2)$.

    \item[(b)] \textit{Propose and accept $(\alpha,h)$:} 
        \begin{itemize}
        \item[(i)] \textit{Propose $\alpha^\dag$:} We transform $\alpha$ to unconstrained parameters $\vartheta = (\mu, \log\{ (1+\phi)/(1-\phi) \}, \log \sigma^2)$ and compute the posterior mode $\hat{\vartheta}$ by maximizing $\pi^*(\vartheta|\gamma, s, y)$. Using the Hessian evaluated at $\hat{\vartheta}$, we construct a Gaussian proposal $N(\hat{\vartheta}, \Sigma_*)$. We draw $\vartheta^\dag$ from this proposal and accept it using a standard MH ratio based on the approximate marginal density $\pi^*(\vartheta|\gamma, s, y)$. If accepted, we map it back to obtain $\alpha^\dag$.
        
        \item[(ii)] \textit{Propose $h^\dag$:} Given the accepted $\alpha^\dag$ and current $s$, we generate $h^\dag$ directly from the approximate conditional density $\pi^*(h | \alpha^\dag, \gamma, s, y)$ using the simulation smoother (\cite{DeShephard(95)}, \cite{DurbinKoopman(02)}).

        \item[(iii)] \textit{MH correction step:} Because the proposal distribution $\pi^*(\alpha, h|\gamma, s, y)$ acts as an analytical proxy for the exact posterior, the final acceptance rate for the joint block $(\alpha^\dag, h^\dag)$ reduces to the ratio of the true likelihood to the approximate mixture likelihood evaluated at the new and old states:
        \begin{align*}
            \min &\left\{ 1, \frac{\pi(\alpha^\dag, h^\dag| \gamma, y)}{\pi(\alpha, h| \gamma,y)} \frac{\pi^*(\alpha, h | \gamma, s, y)}{\pi^*(\alpha^\dag, h^\dag | \gamma, s, y)} \right\} \\
            &= \min \left\{ 1, \frac{\prod_{t=1}^n f(y_t| h_t^\dag, \alpha^\dag,\gamma) \sum_{i=1}^{10} \tilde{p}_{i} g(\tilde{m}_i| h_t, \alpha, s_t=i) }{\prod_{t=1}^n f(y_t| h_t, \alpha,\gamma) \sum_{i=1}^{10} \tilde{p}_{i} g(\tilde{m}_{i}| h_t^\dag, \alpha^\dag, s_t=i) } \right\}.
       \end{align*}
       This lightweight correction ensures that the MCMC algorithm targets the exact posterior distribution without any approximation error.
    \end{itemize}
\end{itemize}

\section{Illustrative numerical examples}
\label{sec:Illustrative examples}

In this section, we evaluate the performance of our proposed Unified Mixture Sampler (UMS) through simulation studies. We consider the SCD model with two different distributions: the Weibull distribution (Weibull-SCD) and the Gamma distribution (Gamma-SCD). For the Gamma-SCD model, the mixture components are adapted by setting $a=2\zeta, b=2 y_t \zeta$, and $c=-1$ in Equations \eqref{equ:mixture p}--\eqref{equ:mixture m}. To demonstrate the versatility of the UMS, we focus on its ability to handle different shape parameters without re-optimization.

We run the MCMC simulation for 50,000 iterations after a 10,000-draw burn-in period.
In all experimental cases, we confirmed that the UMS successfully recovers the true parameter values as shown in Tables  \ref{table:sim_MCMC_weibull} and \ref{table:sim_MCMC_gamma}. The posterior means for the state equation parameters ($\mu, \phi, \sigma$) and the shape parameters ($\gamma, \zeta$) are found to be close to their respective true values, with all 95\% credible intervals covering the true specifications. Furthermore, the inefficiency factors (IFs)\footnote{IF
is calculated by $1 + 2\sum_{s=1}^{\infty}\rho_s$, where $\rho_s$ is the sample autocorrelation at lag $s$. This is interpreted as the ratio of the numerical variance of the posterior mean from the chain to the variance of the posterior mean from hypothetical uncorrelated draws. They are overall small, as expected, which means that the MCMC sampling is close to the uncorrelated sampling.
} for these parameters remained consistently low, indicating high stability and fast convergence of the overall MCMC algorithm. 
The acceptance rates for $(\alpha, h, \gamma)$ are (76.8\%, 96.3\%, 27.7\%) for $\gamma = 0.5$, (77.0\%, 95.9\%, 29.6\%) for $\gamma = 1.0$ in the Webull-SCD model, while those for $(\alpha, h, \zeta)$ are (77.2\%, 95.2\%, 42.7\%) for $\zeta = 1.0$ and  (76.1\%, 89.4\%, 44.1\%) for $\zeta = 2.0$ in the Gamma-SCD model.
\begin{table}[H]
    \footnotesize
      \centering
      \begin{tabular}{llrrrcr}
        \hline 
        $\gamma$ & Param. & True & Mean & Std Dev & 95\% interval & IF \\
        \hline
         & $\mu$ & 0 & 0.215 & 0.268 & (-0.308,  0.775) & 5 \\
         & $\phi$ & 0.97 & 0.959 & 0.014 & ( 0.927,  0.982) & 7 \\
         & $\sigma$ & 0.3 & 0.310 & 0.049 & ( 0.225,  0.416) & 8 \\
        0.5 & $\gamma$ & 0.5 & 0.499 & 0.014 & ( 0.473,  0.527) & 10 \\
         & $h_{100}$ & 2.027 & 1.079 & 0.532 & ( 0.069,  2.15) & 3 \\
         & $h_{500}$ & 1.615 & 0.758 & 0.555 & (-0.307,  1.866) & 3 \\
         & $h_{1000}$ & -0.724 & -0.178 & 0.658 & (-1.42,  1.164) & 2 \\
         \hline
         & $\mu$ & 0 & 0.206 & 0.292 & (-0.358,  0.812) & 23 \\
         & $\phi$ & 0.97 & 0.967 & 0.010 & ( 0.946,  0.984) & 12 \\
         & $\sigma$ & 0.3 & 0.279 & 0.028 & ( 0.228,  0.339) & 9 \\
        1.0 & $\gamma$ & 1.0 & 0.989 & 0.029 & ( 0.934,  1.048) & 8 \\
         & $h_{100}$ & 2.027 & 1.364 & 0.357 & ( 0.691,  2.092) & 3 \\
         & $h_{500}$ & 1.615 & 1.209 & 0.372 & ( 0.507,  1.967) & 3 \\
         & $h_{1000}$ & -0.724 & -0.233 & 0.454 & (-1.065,  0.704) & 3 \\
         \hline
      \end{tabular}
      \caption{Weibull-SCD model. True values, posterior means, posterior standard deviations, 95\% credible intervals, and inefficiency factors. Priors:$\mu \sim N(0, 5^2)$, $\frac{\phi + 1}{2} \sim Beta(1,1)$, $\sigma^2 \sim IG\left( \frac{0.001}{2}, \frac{0.001}{2} \right)$, $\gamma \sim U(0, 10)$.}
      \label{table:sim_MCMC_weibull}
      \normalsize
\end{table}
\begin{table}[H]
    \footnotesize
      \centering
      \begin{tabular}{llrrrcr}
        \hline 
        $\zeta$ & Param. & True & Mean & Std Dev & 95\% interval & IF \\
        \hline
         & $\mu$ & 0 & 0.135 & 0.308 & (-0.47,  0.769) & 12 \\
         & $\phi$ & 0.97 & 0.970 & 0.010 & ( 0.949,  0.987) & 10 \\
         & $\sigma$ & 0.3 & 0.270 & 0.030 & ( 0.217,  0.333) & 10 \\
        1.0 & $\zeta$ & 1 & 1.027 & 0.046 & ( 0.94,  1.121) & 8 \\
         & $h_{100}$ & 2.027 & 1.085 & 0.339 & ( 0.451,  1.779) & 3 \\
         & $h_{500}$ & 1.615 & 1.452 & 0.342 & ( 0.812,  2.151) & 2 \\
         & $h_{1000}$ & -0.724 & -1.008 & 0.526 & (-2.007,  0.059) & 3 \\
         \hline
         & $\mu$ & 0 & 0.124 & 0.324 & (-0.531,  0.771) & 23 \\
         & $\phi$ & 0.97 & 0.971 & 0.009 & ( 0.953,  0.987) & 15 \\
         & $\sigma$ & 0.3 & 0.270 & 0.023 & ( 0.226,  0.318) & 13 \\
        2.0 & $\zeta$ & 2 & 2.038 & 0.104 & ( 1.84,  2.252) & 10 \\
         & $h_{100}$ & 2.027 & 1.267 & 0.299 & ( 0.695,  1.873) & 2 \\
         & $h_{500}$ & 1.615 & 1.763 & 0.296 & ( 1.206,  2.367) & 2 \\
         & $h_{1000}$ & -0.724 & -0.881 & 0.469 & (-1.782,  0.054) & 4 \\
         \hline
      \end{tabular}
      \caption{Gamma-SCD model. True values, posterior means, posterior standard deviations, 95\% credible intervals, and inefficiency factors.       Priors:$\mu \sim N(0, 5^2)$, $\frac{\phi + 1}{2} \sim Beta(1,1)$, $\sigma^2 \sim IG\left( \frac{0.001}{2}, \frac{0.001}{2} \right)$, $\zeta \sim U(0, 10)$.}
      \label{table:sim_MCMC_gamma}
      \normalsize
\end{table}

The primary advantage of the UMS lies in its sampling efficiency for latent variables. We compare the UMS with the single-move slice sampling (SS) method of \cite{MenKolkiewiczWirjanto(15)}. To eliminate the effects of sampling other than the latent states and shape parameters, we assume that ($\mu, \phi, \sigma$) are known and fixed.

Table \ref{table:IFs} presents the inefficiency factors (IFs) for selected latent states ($h_t, t=100,500,1000$), alongside the mean ($\overline{h}$) and the median ($h_{med}$) of the IFs under typical shape parameters ($\gamma = 0.5, 1.0$ and $\zeta = 1.0, 2.0$). The IFs of the UMS are generally less than 7, which are substantially smaller than those resulting from the SS method. 

Furthermore, we compared the computation times required by the two algorithms (Table \ref{table:computation_time}). While the execution time per iteration is slightly longer for the UMS due to the Kalman filter and simulation smoother, its superior sampling efficiency per unit time is evident, particularly when duration clustering is intense (i.e., smaller $\gamma$ or $\zeta$).

\begin{table}[H]
 \footnotesize
 \centering
 \begin{tabular*}{0.8\textwidth}{l@{\extracolsep{\fill}}*{4}{r}}
    \hline 
   Model & MS & SS & MS & SS  \\
     \hline
%     \multicolumn{5}{l}{Weibull-SCD ($\gamma$)} \\
     Weibull-SCD 
     & \multicolumn{2}{c}{$\gamma=0.5$} 
     & \multicolumn{2}{c}{$\gamma=1.0$}\\
     \cline{2-3}   \cline{4-5}  
    \hspace{3mm}$h_{100}$ & 2.4 & 36.0 & 3.0 & 11.3 \\
    \hspace{3mm}$h_{500}$ & 3.6 & 40.8 & 3.1 & 11.5 \\
    \hspace{3mm}$h_{1000}$ & 2.3 & 27.3 & 2.9 & 7.8  \\
    \hspace{3mm}$\overline{h}$ & 6.6 & 81.3 & 5.3 & 19.5 \\
    \hspace{3mm}$h_{med}$ & 5.5 & 67.2 & 4.4 & 19.3 \\
    \hline
%    \multicolumn{5}{l}{Gamma-SCD ($\zeta$)} \\
     Gamma-SCD
     & \multicolumn{2}{c}{$\zeta=1.0$} 
     & \multicolumn{2}{c}{$\zeta=2.0$}\\
     \cline{2-3}   \cline{4-5}  
    \hspace{3mm}$h_{100}$ & 3.2 & 10.2 & 2.3 & 6.8 \\
    \hspace{3mm}$h_{500}$ & 3.0 & 10.6 & 2.4 & 6.8 \\
    \hspace{3mm}$h_{1000}$ & 3.2 & 11.7 & 2.7 & 7.8 \\
    \hspace{3mm}$\overline{h}$ & 5.5 & 25.5 & 4.5 & 15.3\\
    \hspace{3mm}$h_{med}$ & 5.1 & 16.6 & 3.1 & 11.2 \\
    \hline
\end{tabular*}
  \caption{Inefficiency factors of selected $h_t$ and their mean ($\overline{h}$) and median ($h_{med}$). Upper panel presents results for the Weibull-SCD model, and lower panel for the Gamma-SCD model. MS denotes our Unified Mixture Sampler, and SS denotes the Slice Sampler.}
  \label{table:IFs}
\end{table}

\begin{table}[H]
 \footnotesize
 \centering
 \begin{tabular*}{0.8\textwidth}{l@{\extracolsep{\fill}}*{4}{r}}
    \hline 
     & \multicolumn{2}{c}{Weibull-SCD} & \multicolumn{2}{c}{Gamma-SCD} \\
     \cline{2-3} \cline{4-5}
     & $\gamma=0.5$ & $\gamma=1.0$ & $\zeta=1.0$ & $\zeta=2.0$ \\
     \hline
    MS  & 104.2 & 104.5 & 106.7 & 105.0\\
    SS  & 71.7  & 71.7  & 35.4  & 35.8 \\
    \hline
 \end{tabular*}
 \caption{Computational time (seconds) for 50,000 MCMC samples after 10,000 burn-in.}
 \label{table:computation_time}
\end{table}

\section{Conclusion}
\label{sec:Conclusion}

In this paper, we developed the unified mixture sampler (UMS), a versatile MCMC framework for nonlinear non-Gaussian state-space models. By focusing on the widespread `exp-exp' likelihood kernel, we demonstrated that the standard ten-component normal mixture can be dynamically re-centered and rescaled to provide high-precision approximations for a broad class of models. This approach avoids the need for model-specific mixture derivations and provides a common platform for various estimation tasks.

The effectiveness of the UMS was illustrated through its application to the stochastic conditional duration (SCD) model. By dynamically adapting the mixture components to handle unknown shape parameters, such as $\gamma$ in the Weibull distribution or $\zeta$ in the Gamma distribution, we restored the conditionally linear Gaussian structure and utilized efficient simulation smoothers. Simulation experiments across both distributions confirmed that our approach consistently produces inefficiency factors substantially lower than those of the slice sampler. Although the computational time per iteration is higher for the UMS due to the Kalman filter and simulation smoother, it provides significantly more effective samples per unit of time, particularly when duration clustering is intense.

The UMS framework ensures reliable posterior inference through a Metropolis-Hastings correction, making it a robust tool for complex time-varying systems. Given its analytical simplicity and computational speed, this approach serves as an efficient alternative for estimating a wide range of models with positive-valued observations, including logit, Poisson, and various SCD model specifications.

\small
\bibliography{ref_ASVM_paper}	
\normalsize
%\clearpage
%

\end{document}